\documentclass[switch]{article} 

\usepackage{preprint}

\usepackage{amsmath, amsthm, amssymb, amsfonts}

\usepackage[numbers,square]{natbib}
\usepackage[utf8]{inputenc}	
\usepackage[T1]{fontenc}	
\usepackage{xcolor}		
\usepackage[hidelinks]{hyperref}	
\usepackage{booktabs} 		
\usepackage{nicefrac}		
\usepackage{microtype}		
\usepackage{lineno}		
\usepackage{float}			
\usepackage{graphicx}
\usepackage{lipsum}		
\usepackage[bbgreekl]{mathbbol}
\usepackage{tabularx}
\usepackage{subcaption}
\usepackage{mathtools}
\DeclarePairedDelimiter\abs{\lvert}{\rvert}%
\DeclarePairedDelimiter\norm{\lVert}{\rVert}%
\makeatletter
\let\oldabs\abs
\def\abs{\@ifstar{\oldabs}{\oldabs*}}
\let\oldnorm\norm
\def\norm{\@ifstar{\oldnorm}{\oldnorm*}}
\makeatother

\usepackage[ruled,resetcount]{algorithm2e}
\makeatletter
\renewcommand*\l@algocf{\l@figure}
\makeatother
\SetAlgoInsideSkip{smallskip}
\SetAlCapNameFnt{\footnotesize}
\SetAlCapFnt{\footnotesize}

\graphicspath{{Figs/}}

\usepackage{newfloat}
\DeclareFloatingEnvironment[name={Supplementary Figure}]{suppfigure}
\usepackage{sidecap}
\sidecaptionvpos{figure}{c}

\usepackage{titlesec}
\titlespacing\section{0pt}{12pt plus 3pt minus 3pt}{1pt plus 1pt minus 1pt}
\titlespacing\subsection{0pt}{10pt plus 3pt minus 3pt}{1pt plus 1pt minus 1pt}
\titlespacing\subsubsection{0pt}{8pt plus 3pt minus 3pt}{1pt plus 1pt minus 1pt}

\title{Variance and Error in One-Step Phase-Retrieval}

\usepackage{authblk}

\author[1]{Peter J. Christopher}
\author[1]{Timothy D. Wilkinson}
\affil[1]{Centre of Molecular Materials, Photonics and Electronics, University of Cambridge, UK}

\begin{document}

    \maketitle
    
    \begin{abstract}
        Time multiplexed approaches for high frame-rate holographic displays have been around since the invention of One-Step Phase-Retrieval (OSPR) in the early 2000s. When discovered, formulations were created for variance reduction but other image quality metrics were ignored.
        
        This work sets out statistical models for the mean squared error (MSE) and structural similarity index (SSIM) behaviour of OSPR for a range of image types in order to better understand the effect of time multiplexing on visible images. This finds that while observed variances converges to zero as the number of frames per second increases, MSE converges to a non-zero value while SSIM converges quadratically to a non-unitary value. 
    \end{abstract}
    \keywords{Computer Generated Holography  \and One-Step Phase-Retrieval \and Structural Similarity Index} 
    \vspace{0.35cm}

    \normalsize
    \twocolumn
	
	\section{Introduction}
	
    Computer Generated Holography~(CGH) has seen application in projectors and displays since the 80s. In 2004 Cable and Buckley developed a novel approach to real-time hologram generation \cite{cable200453} known as One-Step Phase-Retrieval (OSPR). Unlike many of its predecessors, OSPR relied on time averaging of many time-multiplexed low-quality holograms instead of producing fewer high quality algorithms \cite{buckley2011real, buckley200870, Cable06, Naydenova2011}. The human eye's persistence of vision allows for an assumed linear sum of all frame intensities shown within the previous $\nicefrac{1}{60}$ of a second \cite{kelly1977two, watson1985model, boynton1996linear, adelson1985spatiotemporal}. This is shown in Algorithm~\ref{alg:ospr}. OSPR was later incorporated into much of the work of Light Blue Optics (LBO), and never received the analysis it deserved. 
        
    \begin{algorithm}[htbp]
        \caption{One-Step Phase-Retrieval  \label{alg:ospr}}
        \DontPrintSemicolon
        \KwIn{Target image $T$ and number of sub-frames $N$}
        \KwOut{Output holograms $H[1..N] \leftarrow H'[1..N]$}
        \IncMargin{1em}
        \For{$n \leftarrow 1$ \KwTo $N$}
        {
            \nl Randomise target image phase: $R'_{u,v} = \abs{T_{u,v}}\angle \texttt{Rand} [ 0,2 \pi ]$\;
            \nl Back-propagate the target to the diffraction plane: $H=\mathcal{F}^{-1}\left\{R'\right\}$\;
            \nl Quantise and output the resultant hologram: $H'_n=\texttt{Quantise}\left(H\right)$\;
        }
    \end{algorithm}
    
    The rise of mixed reality systems has seen a resurgence of interest in real-time CGH and in the algorithms required.  In this letter we investigate the statistical properties of the OSPR algorithm shown in Figure~\ref{alg:ospr}. Firstly, relationships for the variance are derived based on earlier work by Cable and Buckley. These are then extended to show the effect on Mean Squared Error (MSE) and Structural Similarity Index (SSIM). The implications on generation performance and framerate is then discussed and recommendations made before conclusions are drawn.
    
    \section{Terminology}
        
    It is assumed that the reader is familiar with the principles of two-dimensional holography. Failing this, the reader is referred to either recent reviews of the topic \cite{doi:10.1080/15980316.2016.1255672, Tsang:18} or to one of the many books on the subject \cite{goodman2005introduction}. In this work we take $x$, $y$ to represent the diffraction field or SLM axes and $u$, $v$ to represent the spatial frequency axes of the replay fields. We take the Discrete Fourier Transform (DFT) between the two to be 
    
    \begin{align}\label{DFT}
    F_{u,v} = \mathcal{F}\{f_{x,y}\}         & = \frac{1}{\sqrt{N_xN_y}}\sum_{x=0}^{N_x-1}\sum_{y=0}^{N_y-1} f_{xy}e^{-2\pi i \left(\frac{u x}{N_x} + \frac{v y}{N_y}\right)}   \\
    f_{x,y} = \mathcal{F}^{ - 1 }\{F_{u,v}\} & = \frac{1}{\sqrt{N_xN_y}}\sum_{u=0}^{N_x-1}\sum_{v=0}^{N_y-1} F_{uv}e^{2\pi i \left(\frac{u x}{N_x} + \frac{v y}{N_y}\right)},
    \end{align}
    
    where the scaling factors are chosen to ensure the conservation of energy inherent in Parseval's Theorem which states that hologram pixels $H_{x,y}$ and replay field pixels $R_{u,v}$ are related by the following relationship
    
    \begin{equation}
    \sum_{x=0}^{N_x-1}\sum_{y=0}^{N_y-1} \abs{H_{x,y}}^2  = \sum_{u=0}^{N_x-1}\sum_{v=0}^{N_y-1} \abs{R_{u,v}}^2
    \end{equation}
    
    \section{OSPR Variance}\label{sec:variance}
    
    In their proceedings paper \cite{cable200453} Cable and Buckley demonstrate that the variance in noise due to $N$ subframes is proportional to the reciprocal of $N$. To demonstrate this we observe that the perceived intensity of a subframe is given as the square of the replay field amplitude $R\overline{R}$ where $R$ is the amplitudes of the replay field given by the fourier transform of hologram $H$. For $N$ subframes the perceived intensity is equal to 
    
    \begin{equation}
    I_{u,v}=R_{u,v}\overline{R_{u,v}}=\frac{1}{N}\sum_{n=1}^{N}\mathcal{F}(H_n)_{u,v}\overline{\mathcal{F}(H_n)_{u,v}}
    \end{equation}
    \begin{equation}
    \forall\quad u \in \mathbb{Z} \cup \left(0, N_x\right] \land v \in \mathbb{Z} \cup \left(0, N_y\right]
    \end{equation}
    
    We assume that the value of $R\overline{R}$ is equal to the target image intensities $T\overline{T}$ plus a noise term $\epsilon'$ with mean $\mu_{\epsilon'}$ and variance $\sigma^2_{\epsilon'}$ where $\epsilon'$ is a complex random variable of circularly symmetric distribution. We also assume that Parseval's theorem holds and that the total energy in the diffraction and replay fields is the same. 
    
    For a single sub-frame system we can write
    
    \begin{equation}
    R_{u,v}\overline{R_{u,v}}=T_{u,v}\overline{T_{u,v}}+\epsilon'_{u,v}.
    \end{equation}
    
    For systems utilising $N$ sub-frames this becomes.
    
    \begin{equation}
    R_{u,v}\overline{R_{u,v}}=T_{u,v}\overline{T_{u,v}}+\frac{1}{N}\sum_{n=1}^{N}\epsilon'_{n,u,v}.
    \end{equation}
    
    Provided $N$ is sufficiently large and $\epsilon$ is independent and identically distributed (i.i.d.) for all $n$, $u$, $v$, the Central Limit Theorem~(CLT) can be applied. Leading to
    
    \begin{equation} \label{combo}
    R_{u,v}\overline{R_{u,v}}=T_{u,v}\overline{T_{u,v}}+\epsilon_{u,v}.
    \end{equation}
    
    where random variable $\epsilon$ is a bivariate Gaussian random variable with mean $\mu_{\epsilon}=\mu_{\epsilon'}$ and variance $\sigma^2_{\epsilon}=\nicefrac{\sigma^2_{\epsilon'}}{N}$. This means that, provided the error terms are independent, the error variance will be equal to the reciprocal of the number of frames.
    
    It is worth noting that the number of sub-frames is anticipated to be low in real world systems and that there will be noticeable non-linearity in the variance due to the CLT.
    
    \section{OSPR Mean Squared Error}\label{sec:mse}
    
    It is tempting to think that, provided that energy conservation is upheld, that the mean squared error $\mu_{\epsilon}=0$ will be equal to the variance. Unfortunately, due to the no this does not follow for reasons we shall discuss individually here. 
    
    The MSE is taken as being the mean difference between reference and target intensities.
    
    \begin{equation} \label{mse}
    Error(T,R) = \frac{1}{N_x N_y}\sum_{x=0}^{x=N_x-1}\sum_{y=0}^{y=N_y-1} \left(R_{u,v}\overline{R_{u,v}} - T_{u,v}\overline{T_{u,v}}\right)^2 
    \end{equation}
    
    As the system is non-linear, the MSE will actually consist of a variance term summed with a constant bias term. Two common cases cause the bias
    
    \subsection{Conjugate Symmetry}
    
    The first reason for bias is due to the conjugate image symmetry requirements in binary devices which mandates that a replay field must be equal to itself when subject to a $180^{\circ}$ rotation around the centre of the field. i.e. for a pixel $R_{u,v}$ on a replay field of size $N_x$, $N_y$ the replay field values will follow the relationship
    
    \begin{equation} 
    R_{u,v} \equiv R_{N_x-u,N_y-v}\quad\forall\quad u \in \mathbb{Z} \cup \left(0, N_x\right] \land v \in \mathbb{Z} \cup \left(0, N_y\right].
    \end{equation}
    
    This symmetry relationship fails the i.i.d. requirements. Instead of Eq.~\ref{combo} we must write
    
    \begin{align}
    R_{u,v}\overline{R_{u,v}}&=R_{N_x-u,N_y-v}\overline{R_{N_x-u,N_y-v}}\nonumber \\ 
    &=\frac{T_{u,v}\overline{T_{u,v}}+\epsilon_{u,v}+T_{N_x-u,N_y-v}\overline{T_{N_x-u,N_y-v}}+\epsilon_{N_x-u,N_y-v}}{2}
    \end{align}
    
    which leads to an equation for the bias $\text{Bias}_{cs}(T,R)$ due to conjugate symmetry of
    
    \begin{align}
    &\text{Bias}_{cs}(T,R)^2  = \nonumber \\ 
    &\frac{1}{N_x N_y}\sum_{u=0}^{N_x-1}\sum_{v=0}^{N_y-1} \left(\frac{\abs{R_{u,v}\overline{R_{u,v}}-R_{N_x-u,N_y-v}\overline{R_{N_x-u,N_y-v}}}}{2}\right)
    \end{align}
    
    For rotationally symmetric target images, this is expected to be zero.
    
    While this symmetry requirement is well understood, more involved symmetry requirements exist for almost every binary device. For example, a binary phase device not modulating to an interval of $\pi$ will introduce complicated periodic symmetry requirements.
    
    \subsection{Intensity Distribution}
    
    The second source of bias is due to the nature of the noise distribution. While the noise variance converges to zero as $N\rightarrow \infty$, the intensity is determined by the square of the replay amplitudes. This translates to the mean of the image intensities not being linearly related to the mean of the image amplitudes.
    
    If $\epsilon$ is taken as being circularly symmetric and normally distributed then Eq.~\ref{combo} has a Rician PDF
    
    \begin{align}
    p_{R\overline{R}}(R_{u,v}\overline{R_{u,v}}) = & \frac{R_{u,v}\overline{R_{u,v}}}{\sigma_{\epsilon'}^2}\exp\left(-\frac{(R_{u,v}\overline{R_{u,v}})^2+(T_{u,v}\overline{T_{u,v}})^2)}{2\sigma_{\epsilon'}^2}\right)  \nonumber \\
    & J_0\left(\frac{R_{u,v}\overline{R_{u,v}}T_{u,v}\overline{T_{u,v}}}{\sigma_{\epsilon'}^2}\right),
    \end{align}
    
    where $J_0$ is the first Bessel function. In the limit as $\nicefrac{T_{u,v}\overline{T_{u,v}}}{\sigma_{\epsilon'}} \rightarrow \infty$, this tends to the Gaussian distribution while as $\nicefrac{T_{u,v}\overline{T_{u,v}}}{\sigma_{\epsilon'}} \rightarrow 0$ this tends to the Rayleigh distribution.
    
    If we take the target magnitudes as having an amplitude distribution of $p_{T\overline{T}}(T_{u,v}\overline{T_{u,v}})$ then can write an integral for the bias $\text{Bias}_{id}(T,R)$ due to the intensity distribution of
    
    \begin{align}\label{mseeqn}
    \text{Bias}_{id}(T,R)^2  =& \int_{0}^{\infty}p_{T\overline{T}}(T_{u,v}\overline{T_{u,v}})\int_{0}^{\infty} p_{R\overline{R}}(R_{u,v}\overline{R_{u,v}})  \\
    &\left(R_{u,v}\overline{R_{u,v}} - T_{u,v}\overline{T_{u,v}}\right)^2 d(R_{u,v}\overline{R_{u,v}})d(T_{u,v}\overline{T_{u,v}}) \nonumber
    \end{align}
    
    Analytical solutions are possible but complex. Numerical solutions are practical however. The key observation is that while variance $\sigma_{\epsilon}$ of perceived intensity over $N$ subframes decreases as the reciprocal of $N$ the mean $\mu^2_{\text{id},\epsilon}$ of perceived intensity over $N$ subframes remains a function of $\sigma_{\epsilon'}$. 
    
    \begin{figure}[h]
        \centering
        {\includegraphics[trim={0 0 0 0},width=0.8\linewidth,page=1]{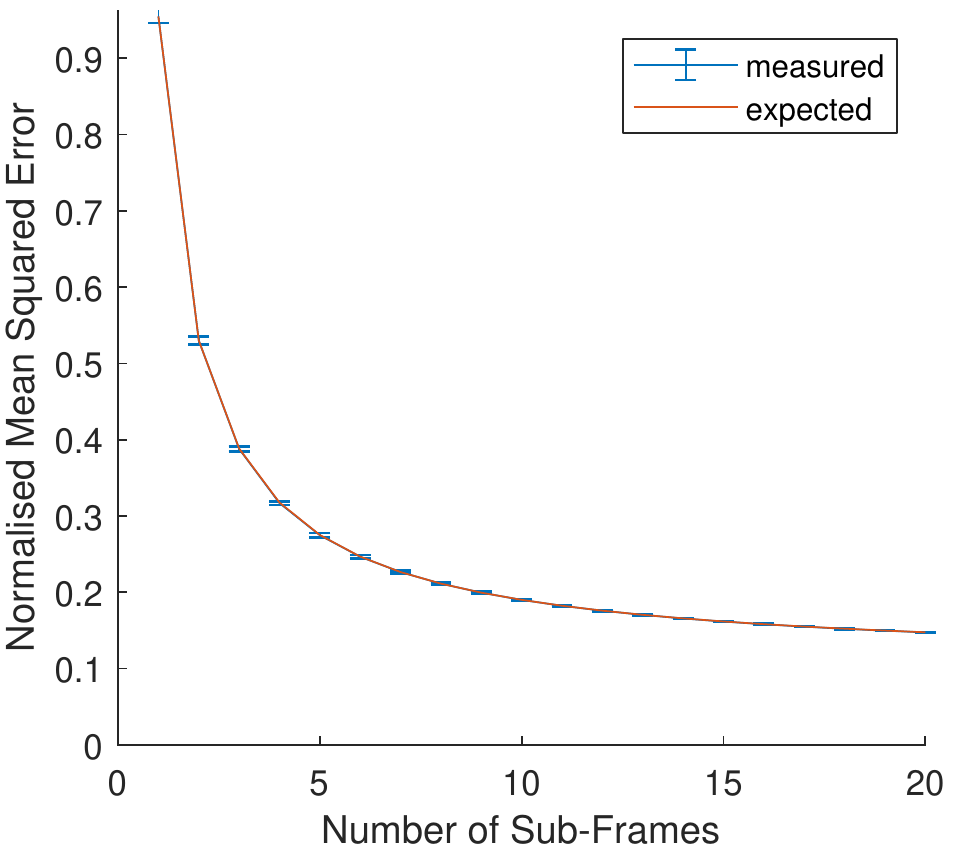}}
        \caption{Time averaged mean squared errors for OSPR. Values are taken as being the mean of $100$ independent runs with error bars showing two standard deviations. The $512\times512$ pixel \textit{Mandrill} test image has an artificially induced symmetry and is modelled for a binary phase SLM.}
        \label{fig:ospr_001}
    \end{figure}
    
    \begin{figure}[h]
        \centering
        {\includegraphics[trim={0 0 0 0},width=0.6\linewidth,page=1]{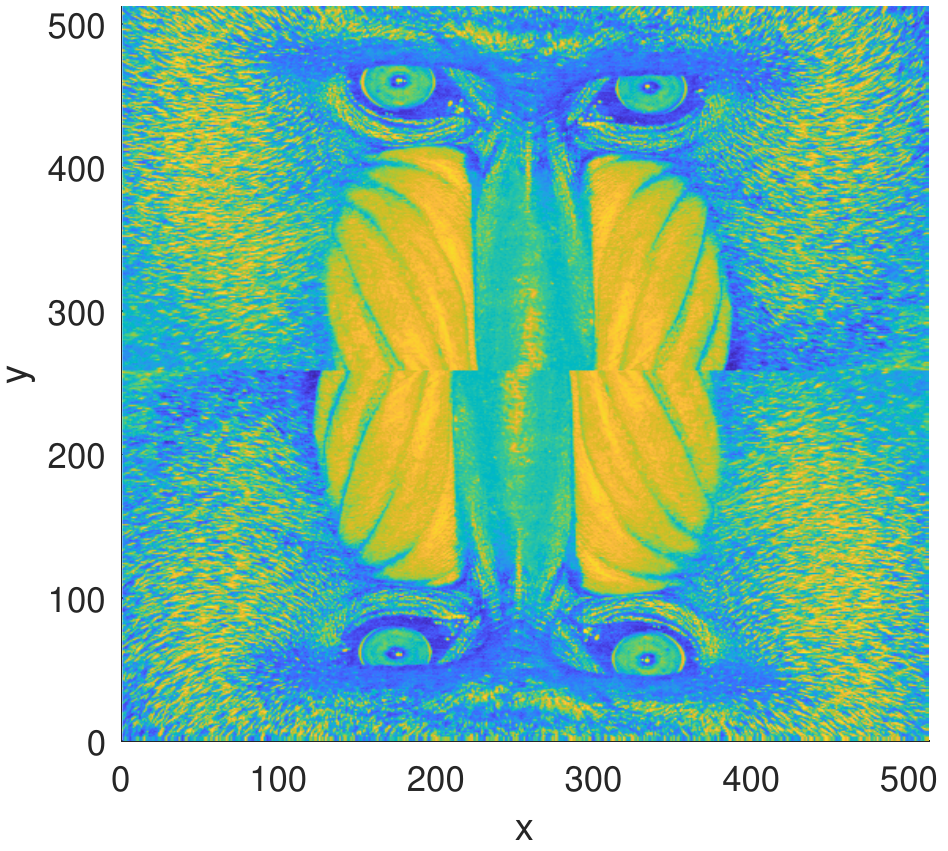}}
        \caption{The $512\times512$ pixel \textit{Mandrill} test image used with artificially induced symmetry.}
        \label{fig:ospr_001b}
    \end{figure}
    
    In practice regression may be used. For example Figure~\ref{fig:ospr_001} shows the convergence of the MSE against number of subframes for a modified \textit{Mandrill} test image which provides a close fit to the expected $E=A+\nicefrac{B}{N}$ curve. In Figure~\ref{fig:ospr_001} the bias $\text{Bias}_{id}(T,R)$ is equal to $0.068$ with initial variance of $\sigma^2_{\epsilon'}$ equal to $0.884$.
    
    \begin{equation}\label{mseref}
    Error(T,R) = \text{Bias}_{id}(T,R)^2 + \sigma_{\epsilon}^2 
    \end{equation}
    
    As both $\text{Bias}_{id}(T,R)^2$ and $\sigma^2_{\epsilon}$ are determined purely by the intensity distribution for images i.i.d. in $n$, $u$ and $v$ this convergence graph will be equal for similarly distributed images. An interesting observation of this is that uniformly distributed image magnitudes will have a higher bias than Gaussian distributions
    
    \subsection{Performance}
    
    Table~\ref{distributions} shows examples of the bias and variance for selected amplitude distributions. As expected from our earlier observations, amplitude distributions with more terms near 0 had a larger bias term than those with amplitude terms with fewer terms near from 0. In the most extreme case, constant amplitude, the bias term was negligible. Of especial interest is that natural images such as \textit{Mandrill} and \textit{Peppers} which have a more central distribution had a smaller bias than uniformly distributed images.
    
    \begin{table}
        \caption{\label{distributions} Mean squared error parameters from Eq.~\ref{mseref} for different amplitude distributions and uniform  $\left[-\pi,\pi\right)$ phase distribution.}
        \begin{tabular}{ccccc}
            \toprule
            Amplitude     & \multicolumn{2}{c}{Measured MSE} & \multicolumn{2}{c}{Simulated MSE}  \\
            Distribution  & $\text{Bias}_{id}(T,R)$ & $\sigma^2_{\epsilon'}$  & $\text{Bias}_{id}(T,R)$ & $\sigma^2_{\epsilon'}$  \\
            \midrule
            Uniform   & $0.452 $ & $ 0.797$ & $0.455$ & $ 0.794$  \\
            Constant  & $0.001$ & $ 0.799$ & $0.000$ & $ 0.797$ \\
            Mandrill  & $0.068$ & $ 0.884$ & $0.067$ & $ 0.883$ \\
            Peppers   & $0.069$ & $ 0.883$ & $0.066$ & $ 0.884$ \\
            \bottomrule
        \end{tabular}
    \end{table}

    Table~\ref{distributions} also shows \textit{measured} and \textit{simulated} values. Measured values were determined by taking the mean and deviations of running OSPR for a given number of subframes. Simulated values were taken by numerically integrating Eq.~\ref{mseeqn} for the given distribution of amplitudes. The simulated values were accurate to within $1\%$ of the measured values.

    \section{OSPR Structural Similarity Index}\label{sec:ssim}
    
    In Section~\ref{sec:variance} we reported on the relationship between variance and number of subframes for the OSPR algorithm In Section~\ref{sec:mse} we developed a similar relationship for the bias term given in Eq.~\ref{mseeqn}. From this we were able to determine the expected MSE for a given distribution of amplitudes.
    
    For displays viewed by the human eye, the Structural Similarity Index~(SSIM) is more commonly used than MSE as it has been shown to more closely correspond to ocular visual quality.\cite{wang2004image} Unlike MSE where the pixel errors are spatially independent, SSIM is determined from moving two $8\times 8$ pixel windows $T$ and $R$ across the target and reconstruction images.
    
    \begin{equation}\label{ssim}
    SSIM(T,R) = \underbrace{\frac{\left(2\mu_T\mu_R+c_1\right)} {\left(\mu_T^2+\mu_R^2+c_1\right)}}_{S_1}\underbrace{\frac{\left(2\sigma_{TR}+c_2\right)} {\left(\sigma_T^2+\sigma_R^2+c_2\right)}}_{S_2}
    \end{equation}
    
    where $\mu_T$ and $\mu_R$ are the window means; $\sigma_T$ and $\sigma_R$ are the window variances; $\sigma_{TR}$ is the covariance of the two window and $c_1$ and $c_2$ are functions of pixel dynamic range, $L$, where $c_1=(k_1L)^2$ and $c_2=(k_2L)^2$. $k_1$ and $k_2$ are usually taken as $0.01$ and $0.03$ respectively. Pixel dynamic range can be taken as being the total number of possible target states.
    
    \begin{figure}[h]
        \centering
        {\includegraphics[trim={0 0 0 0},width=\linewidth,page=1]{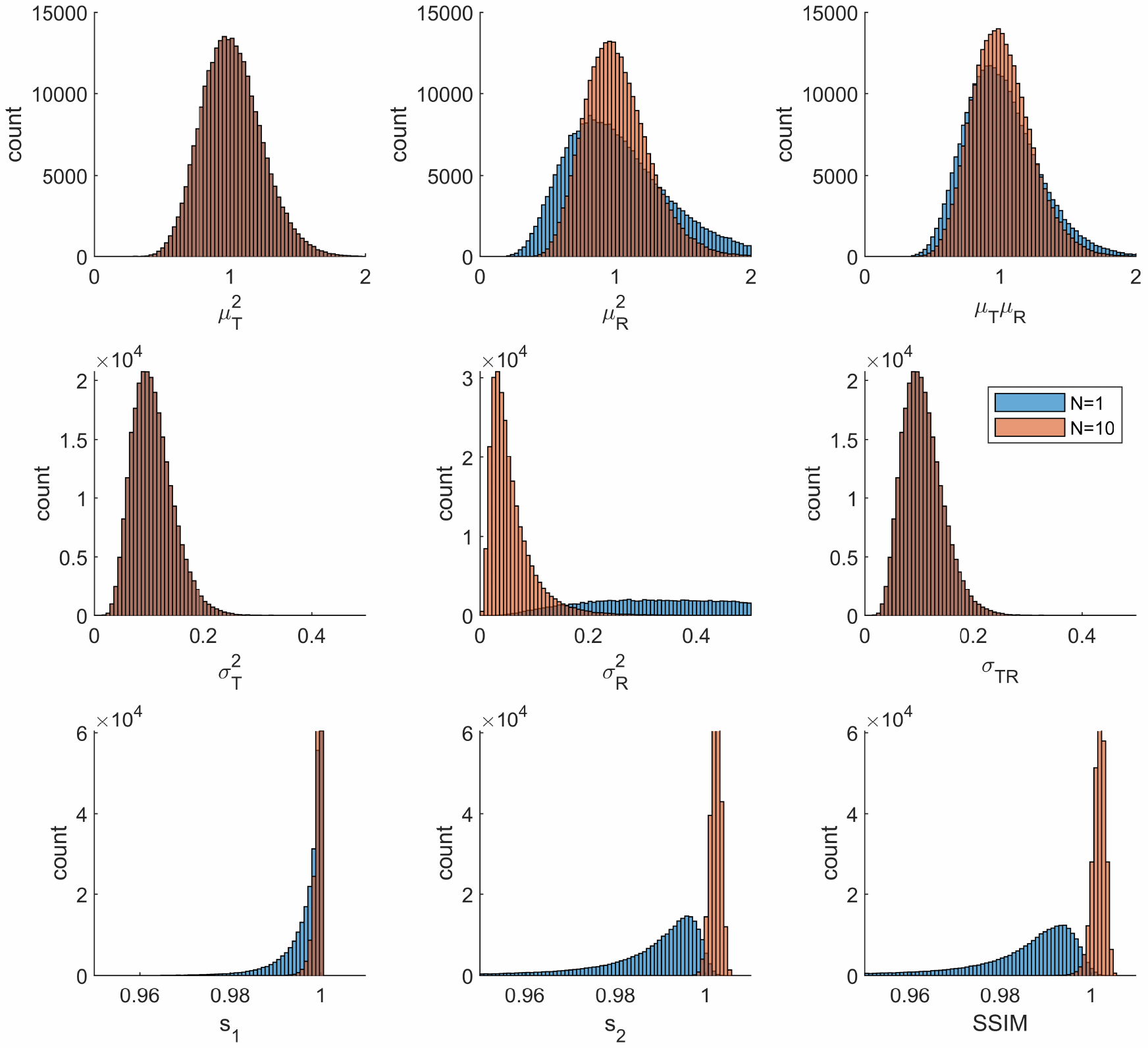}}
        \caption{Evolution of components of the SSIM equation over time. Image used is a $512\times512$ uniformly distributed amplitude.}
        \label{fig:ospr_004}
    \end{figure}
    
    Figure~\ref{fig:ospr_004} shows the components of this relationship for a uniformly distributed amplitude image for 1 and 10 subframes. As expected $\mu_T^2$ and $\sigma_T^2$ do not change with time. The mean value for $\mu_R^2$ stays as 1 but the summing effect of more subframes means that the distribution becomes narrower. The mean value for $\sigma_R^2$ follows the expected $\nicefrac{1}{N}$ relationship and can be assumed identical to $\sigma^2_{\epsilon'}$ for large windows.
    
    It can also be seen from Figure~\ref{fig:ospr_004} that $s_2$ is more prominent than $s_1$ as a factor. If we set $s_1=1$ then we obtain
    
    \begin{equation}\label{ssim2}
    SSIM(T,R) \approx \frac{c_2}{\left(2\sigma_T^2+\nicefrac{\sigma^2_{\epsilon'}}{N}+c_2\right)}
    \end{equation}
    
    \begin{figure}[h]
        \centering
        {\includegraphics[trim={0 0 0 0},width=\linewidth,page=1]{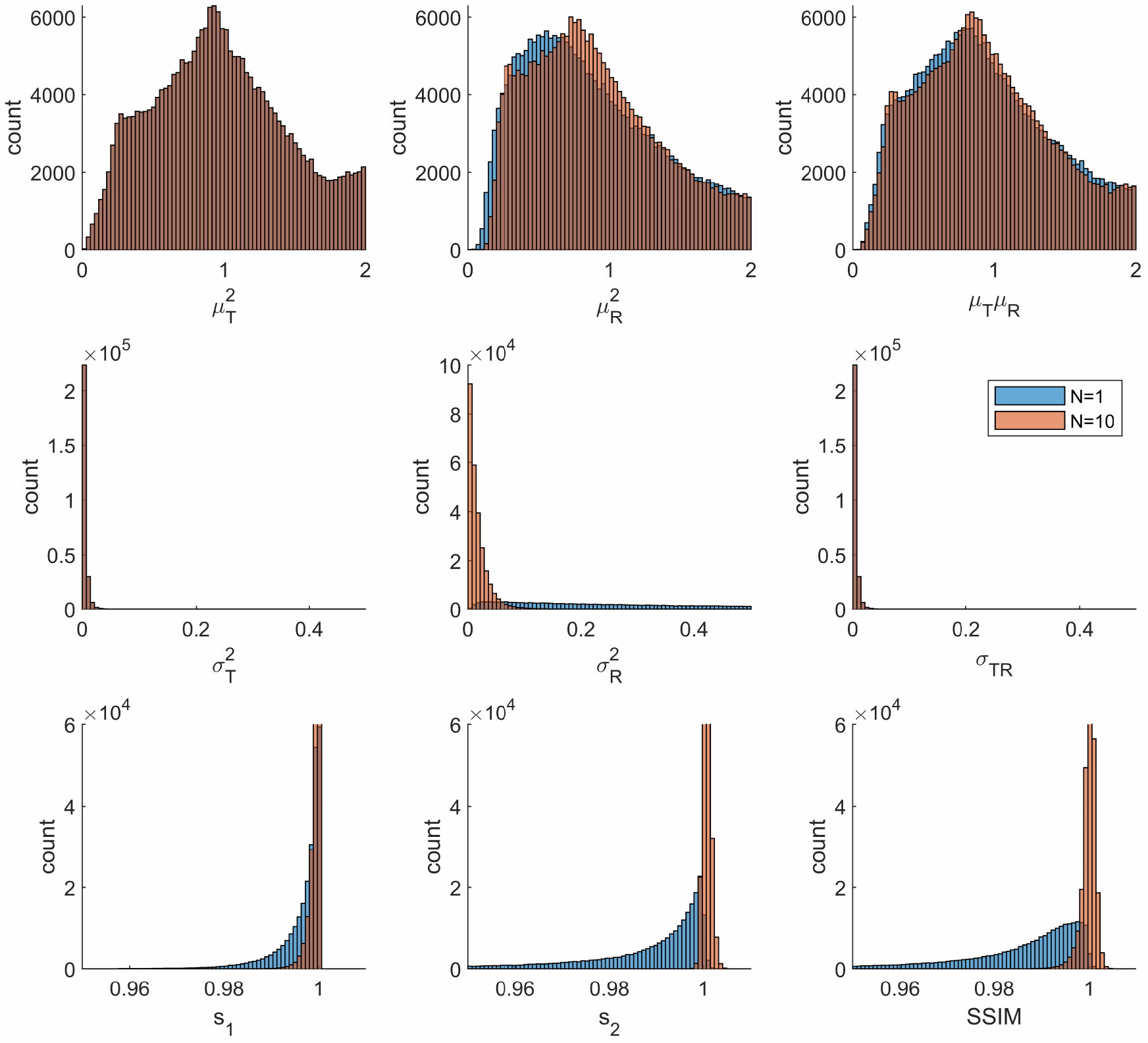}}
        \caption{Evolution of components of the SSIM equation over time. Image used is the $512\times512$ \textit{Mandrill} test image.}
        \label{fig:ospr_004b}
    \end{figure}
    
    Figure~\ref{fig:ospr_004b} shows the same components for the \textit{Mandrill} test image. While the distributions have changed, the behaviour and properties can be seen to be similar. 
    
    \begin{figure}[h]
        \centering
        {\includegraphics[trim={0 0 0 0},width=0.8\linewidth,page=1]{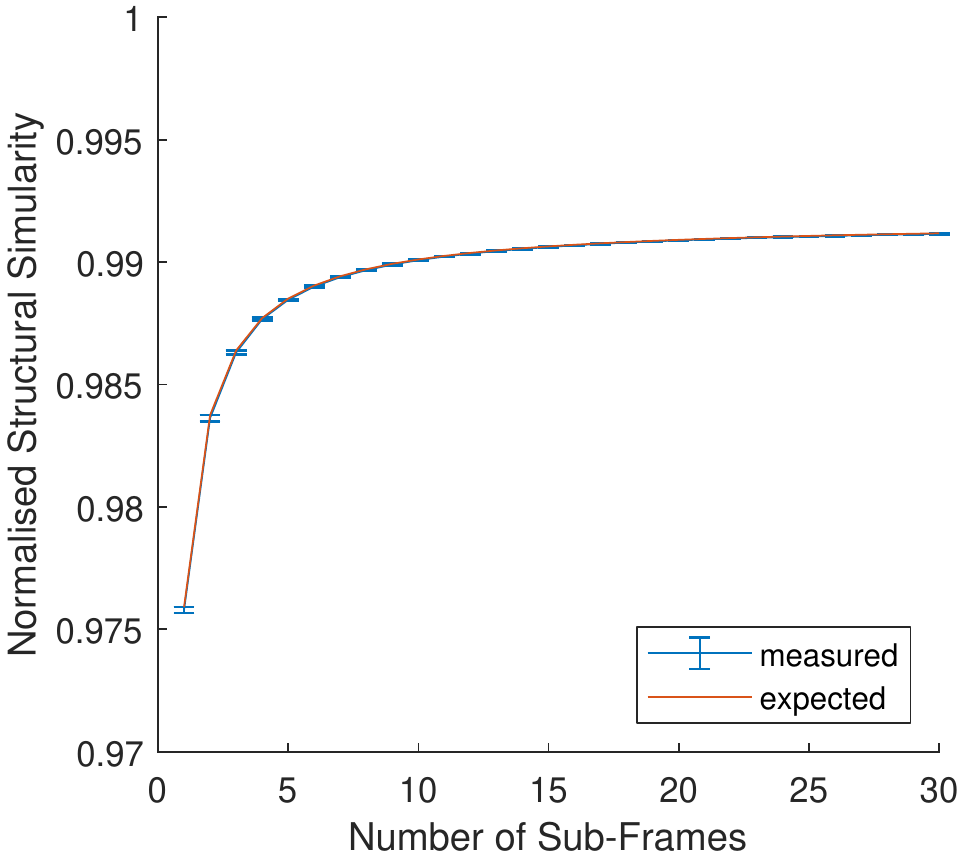}}
        \caption{Time averaged structural similarity for OSPR. Values are taken as being the mean of $100$ independent runs with error bars showing two standard deviations.  The $512\times512$ pixel \textit{Mandrill} test image has an artificially induced symmetry and is modelled for a binary phase SLM.}
        \label{fig:ospr_002}
    \end{figure}

    By integrating Eq.~\ref{ssim2} numerically we can obtain the result shown in Figure~\ref{fig:ospr_002} which provides a close fit to the measured behaviour. 
    
    \section{Conclusion}
        
    In this work we have briefly reintroduced the OSPR algorithm. We then summarised the original analysis made on intensity variance reduction with sub-frame count. We then proceeded to discuss the case of MSE and we presented a numerical relationship for estimating the bias and error against number of sub-frames. This was found to provide accurate estimations of expected MSE for given amplitude distribution with prediction error within $1\%$. 
    
    This was then extended to cover the case of SSIM improvement against sub-frame. This analysis again provided accurate estimates of convergence.
    
    Unsurprisingly, though variance converges to zero, MSE does not converge to zero and SSIM does not converge to unity. This is due to the non-linear square relationship between amplitude and intensity. Despite this, the improvement in MSE and SSIM is significant and OSPR remains a viable algorithm for real-time holography.

	\bibliography{references}
	
\end{document}